# AN UNSUPERVISED LEARNING-BASED FRAMEWORK FOR EFFECTIVE REPRESENTATION EXTRACTION OF REACTOR ACCIDENTS

**Chengyuan Li**
Science and Technology on Reactor System Design Technology Laboratory, Nuclear Power Institute of China
Chengdu, Sichuan, China

**Meifu Li**
Science and Technology on Reactor System Design Technology Laboratory, Nuclear Power Institute of China
Chengdu, Sichuan, China

**Zhifang Qiu**
Science and Technology on Reactor System Design Technology Laboratory, Nuclear Power Institute of China
Chengdu, Sichuan, China

ABSTRACT: Compared to knowledge-based diagnostic systems, data-based methods tend to perform better in terms of speed and accuracy in diagnosing reactor accidents, and have significant advantages in terms of scalability of models. With the increasing use of high-precision system analysis programs in nuclear engineering, the number of high-fidelity computational data for accident simulation is exploding. Therefore, an algorithm that can achieve both automatic extraction of low-dimensional features from the data and guarantee the validity of the features is needed to improve the performance and confidence of the accident diagnosis system. This study proposes an autoencoder-based autonomous learning framework, namely Padded Auto-Encoder (PAE), which is able to automatically encode accident monitoring data that has been noise-added and with partially missing data into low-dimensional feature vectors via a Vision Transformer-based encoder, and to decode the feature vectors into noise-free and complete reconstructed monitoring data. Thus, the encoder part of the framework is able to automatically infer valid representations from partially missing and noisy monitoring data that reflect the complete and noise-free original data, and the representation vectors can be used for downstream tasks for accident diagnosis or else. In this paper, LOCA of HPR1000 was used as the study object, and the PAE was trained by an unsupervised method using cases with different break locations and sizes as the dataset. The encoder part of the pre-trained PAE was subsequently used as the feature extractor for the monitoring data, and several basic statistical learning algorithms for predicting the break locations and sizes. The results of the study show that the pre-trained diagnostic model with two stages has a better performance in break location and size diagnostic capability with an improvement of 41.62% and 80.86% in the metrics respectively, compared to the diagnostic model with end-to-end model structure.

## 1. INTRODUCTION

The reactor accident diagnostic system plays an essential role in the accident management process, as it largely influences which accident mitigation measures should be taken by the operator to bring the reactor from abnormal condition to a safe state.

Since data-based accident diagnosis systems have strong advantages in scalability and speed of inference, an increasing number of diagnostic systems are being built using deep learning (DL) techniques[1]. However, while DL-based diagnostics can automatically capture features of the raw data through deep non-linear mapping to enable an end-to-end inference approach, this black-box model is very prone to overfitting the network by capturing unnecessary features and can make unexpectedly severe inference errors for small input disturbances such as generation of adversarial signals or missing information, which can be fatal in reactor accident diagnosis can be fatal. To overcome these problems in DL-based diagnostic systems, additional means are needed to ensure the validity of the raw data features extracted by the model, and the most effective means currently available is the use of autoencoder (AE) algorithms for feature extraction. Li et al. [2] proposed a method for feature extraction and clustering of transients, using CNN as the backbone network of AE. Kim et al. [3] used Variational Auto Encoder to extract transient features and perform anonymous recognition. NAITO et al. [4] proposed a two-stage AE for feature extraction of detection parameters within a time window to achieve abnormality recognition. However, problems with the method applied by previous researchers remain as follows:

First, the robustness of feature extraction is hardly analyzed for different levels of noise. On the one hand, noise needs to be added as the data applied for model training is usually based on full-stack simulators or system analysis programs, which are smoother compared to real scenarios; on the other hand, the noise level of the detection parameters changes significantly under different electromagnetic noise disturbances, so noise of different levels needs to be added.

Secondly, the effect of missing data on feature extraction is unconsidered. As there is a degree of probability of failure of reactor monitoring instrumentation during anomalous reactor transients, the effectiveness of feature extraction in the absence of some instrumentation data needs to be accounted for.

Finally, the low-dimensional representation of operational states relies on a large scale of operational monitoring parameters and requires a model backbone network with a global view and capable of high-speed computation for feature extraction. However, previous work has focused more on the downstream tasks after feature extraction, with little discussion of the theory of feature extraction model construction.

Based on the above insight, inspired by the Vision Transformer, which is entirely based on a self-attentive mechanism in machine vision tasks [5],and its good performance in the Masked Auto-Encoder task [6],a network structure named Padded Auto-Encoder for reactor monitoring parameter feature extraction is proposed, taking into account the actual needs in reactor accident diagnosis. This method first patches the monitoring parameters that have been sampled over time with noise of varying signal-to-noise ratios; then randomly zeroes the data within the patches in a certain proportion; subsequently adds the sequence information of the patches that can be retained by gradient descent through parameter learning, and appends a learnable class token to the head of the patch sequence. In the encoder section, the sequence passes through several Transformer blocks containing multiple self-attentive layers and fully connected residuals, and then undergoes an LSTM encoding towards the head of the patch sequence, where the obtained state variables $c_t$ are connected by feedforward to get the encoded vector. In the decoder stage, the encoded vector is mapped into the patch sequence dimension through a fully connected network, and after several layers of Transformer is then de-patched and reduced to the dimension of the original input monitoring parameters

The contributions of this paper are presented as follows:
1. An idea is proposed to treat reactor monitoring data as a special two-dimensional image with time-series features and to capture the global effective features of monitoring parameters in a parallel manner using the Vision Transformer with self-attention mechanism;
2. A padded autoencoder model is proposed that not only reduces the noise of monitoring data with different signal-to-noise ratios, but also automatically completes monitoring data with up to 40% random missing;
3. For diagnostic tasks for LOCA accidents, the downstream diagnostic model utilizing statistical learning algorithms with the features extracted by the encoder from monitoring data as input outperformed several other end-to-end diagnostic models

The paper is organized as follows: we first introduce our proposed model in Section 2; in Section 3, we detail the experimental methodology and discuss the experimental results; and finally, in Section 4, we conclude the text and provide an outlook for future research.

## 2. PROPOSED METHOD

This paper begins with a brief introduction to the task. Reactor accident transients are caused by an initiating event and are reflected in a number of thermophysical monitoring parameters. The task of reactor accident diagnosis is to determine the type and severity of the initiating event based on the changes in the monitored parameters. As each parameter $\boldsymbol{u}_d$ is sampled as a discrete sequence of analogue signals, the monitoring data over a period of time is a two-dimensional tensor $\mathcal{X}$ made up of multiple vectors, and the ordering of the vector elements contains time series information.

$$u_d(t) = \sum_{n=1}^{\infty} u(nT)\delta(t-nT)$$
$$\boldsymbol{u}_d(n) = u_d(nT) \in \mathbb{R}^{p \times 1}, n \in \{1,2,\ldots,p\} \quad (1)$$
$$\mathcal{X} = [\boldsymbol{u}_{d,1}; \boldsymbol{u}_{d,2}; \ldots; \boldsymbol{u}_{d,l}] \in \mathbb{R}^{p \times l}$$

where $T$ is the sampling period; $p$ is the number of samples; and $l$ is the number of monitoring parameters. The tensor $\mathcal{X}$ contains tens of thousands of elements, so if a low-dimensional vector $\mathcal{X} \rightarrow \boldsymbol{latent} \in \mathbb{R}^d$ can be used to characterize the monitoring data over this period of time and satisfy $d \ll p \times l$, the low-dimensional vector can be used directly for further identification of anomalous transients or for other types of downstream tasks.

The overall framework of our approach is shown in Figure 1, which is an encoder-decoder type of feature extraction process. First, we add Gaussian noise with different signal-to-noise levels for each of the parameters $\mathcal{X}_n = \text{Noise}(\mathcal{X}; snr)$. The parameters are then patched at the parameter level.

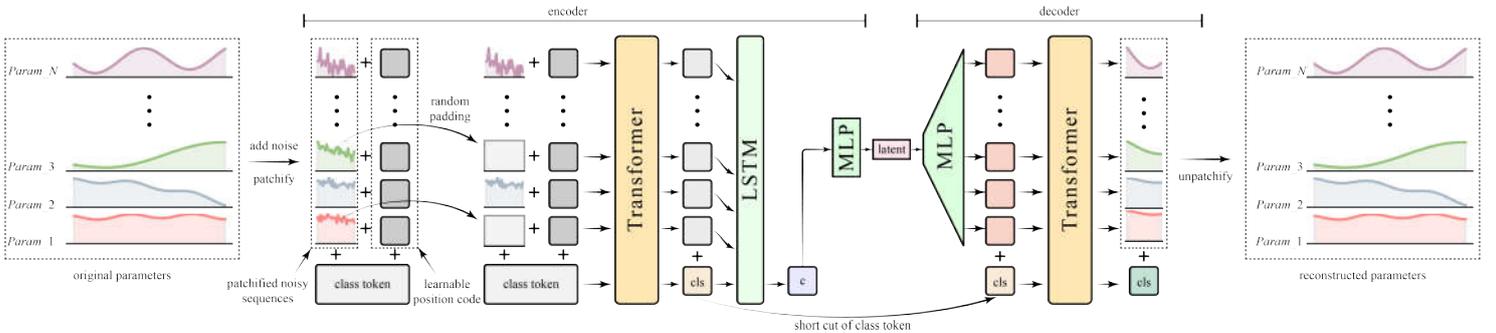

Figure 1 Structural diagram of Padded Auto-Encoder

$$\left[ patch_{i,1}; patch_{i,2}; \ldots; patch_{i,m} \right] = \boldsymbol{u}_{d,i} \quad (2)$$
$$patch_{i,m} \in \mathbb{R}^{(p/m) \times 1}$$

where $i$ indicates the serial number of the monitoring parameter. Each patch contains a segment of local variation of the monitoring parameters, so that the original two-dimensional tensor is transformed into a sequence of word vectors with the patch size as the vector dimension.

$$\mathcal{X}_p = \sum_{i=1}^{l} \sum_{j=1}^{m} \boldsymbol{e}_i \otimes \boldsymbol{e}_j \otimes patch_{i,j}^{\top}$$
$$= \begin{bmatrix} \left[ patch_{1,1}; patch_{1,2}; \cdots; patch_{1,m} \right]^{\top} \\ \left[ patch_{2,1}; patch_{2,2}; \cdots; patch_{2,m} \right]^{\top} \\ \cdots \quad \cdots \quad \cdots \\ \left[ patch_{l,1}; patch_{l,2}; \cdots; patch_{l,m} \right]^{\top} \end{bmatrix} \quad (3)$$
$$\in \mathbb{R}^{(l \cdot m) \times (p/m)} \xrightarrow[D=p/m]{N=l \cdot m} \mathbb{R}^{N \times D}$$

where $\boldsymbol{e}_i$ denotes the unit vector with the $i$-th element being 1; and $\otimes$ is the Kronecker product operator. To improve the robustness of feature extraction against partially missing data, the word vector sequence is masked by a randomly selected proportion, i.e. the data in that patch region is zeroed out. The word vector sequence is added with a layer of positional encoding containing learnable parameters $\boldsymbol{E}_{pos} \in \mathbb{R}^{(N+1) \times D}$, to ensure that the sequence's sequential information is retained. A learnable parameter of the same size as the patch $\boldsymbol{x}_{class} \in \mathbb{R}^{1 \times D}$ is added to the header of the sequence to enable the representation of the class information of the word vector sequence through multi-layer encoding.

$$\mathcal{X}_0 = \left[ \boldsymbol{x}_{class}; \mathcal{X}_p^{\top} \right]^{\top} + \boldsymbol{E}_{pos} \in \mathbb{R}^{(N+1) \times D} \quad (4)$$

Subsequently, in the encoder stage, the sequence of word vectors is encoded through multiple layers of Transformers containing multi-headed self-attention, capturing remotely relevant information between patch blocks through a highly parallelised process. For each layer of Transformer passed through, the encoding is done as follows.

$$\mathcal{X}_q' = \mathrm{MSA}\left(\mathrm{LN}\left(\mathcal{X}_{q-1}\right)\right) + \mathcal{X}_{q-1}$$
$$\mathcal{X}_q = \mathrm{MLP}\left(\mathrm{LN}\left(\mathcal{X}_q'\right)\right) + \mathcal{X}_q', \ q \in \{1, 2, \cdots, Q\} \quad (5)$$

where $Q$ is the number of Transformer layers passed; MSA is the Multi-headed Self-Attention calculation; LN is the Layer Normalization calculation; and MLP is the Multi-Layer Perceptron. The computation process of the multi-headed self-attentive mechanism is as follows.

$$[\boldsymbol{q}, \boldsymbol{k}, \boldsymbol{v}] = \mathcal{X} \boldsymbol{U}_{qkv}, \ \mathcal{X} \in \mathbb{R}^{N \times D}, \boldsymbol{U}_{qkv} \in \mathbb{R}^{D \times 3D_h}$$
$$\boldsymbol{A} = \mathrm{softmax}\left( \boldsymbol{q}\boldsymbol{k}^{\top} / \sqrt{D_h} \right)$$
$$\boldsymbol{SA} = \boldsymbol{A}\boldsymbol{v} \quad (6)$$
$$\boldsymbol{MSA} = \left( \sum_{i=1}^{k} \boldsymbol{e}_i \otimes \boldsymbol{SA}_i(\mathcal{X}) \right) \boldsymbol{U}_{msa} \in \mathbb{R}^{N \times D}$$

where $D_h$ is the dimensionality of the word vector mapping; and $\boldsymbol{U}_{msa} \in \mathbb{R}^{(k \cdot D_h) \times D}$ is the multi-headed self-attentive dimensionality transformation matrix. As the data after Transformer encoding is still a sequence of word vectors of the same length, a layer of LSTM is used to encode the sequence in order to compress the encoded vector sequence.

$$\begin{cases} i_t = \sigma\left( W_{ii} X_t + b_{ii} + W_{hi} h_{t-1} + b_{hi} \right) \\ f_t = \sigma\left( W_{if} X_t + b_{if} + W_{hf} h_{t-1} + b_{hf} \right) \\ g_t = \tanh\left( W_{ig} X_t + b_{ig} + W_{hg} h_{t-1} + b_{hg} \right) \\ O_t = \sigma\left( W_{io} X_t + b_{io} + W_{ho} h_{t-1} + b_{ho} \right) \\ c_t = f_t \odot c_{t-1} + i_t \odot g_t \\ h_t = O_t \odot \tanh(c_t) \\ t \in \{1, 2, \cdots, N\} \end{cases} \quad (7)$$

where $N$ is the length of this vector sequence; $W, b$ are the weights and biases to be learned; $X_N = \boldsymbol{x}_{class}$. The LSTM kernel moves in the direction of scanning from the tail of the sequence to the class token, due to the fact that the state variables within the LSTM retain the most information about the last input data. The state variables of the LSTM are then projected through a three-layer fully connected network to the lower dimensional features, i.e., $\boldsymbol{latent} = \mathrm{MLP}(c_N)$.

In the decoder stage, the hidden variables are first transformed directly through the fully connected network to the same dimension of the word vector sequence as in the encoder stage $\mathcal{X}_{de,0}' = \mathrm{MLP}(\boldsymbol{latent}) \in \mathbb{R}^{N \times D}$. At the same time, the class token in the encoder stage is directly copied to the head of the initial word vector sequence in the decoder stage $\mathcal{X}_{de,0} = \left[ \boldsymbol{x}_{class}; \mathcal{X}_{de,0}' \right] \in \mathbb{R}^{(N+1) \times D}$, which can include sufficiently rich category information in the decoder's word vector sequence. The word vector sequence is then encoded by a multi-layer Transformers. The encoded result is transformed to the original monitoring parameter dimension by an inverse patching, i.e., a two-dimensional tensor $\mathcal{X}_{re}$ is reconstructed.

After a number of training episodes, if the autoencoder is able to not only reconstruct the missing data but also filter out the artificially added noise to some extent, the model can be considered to be able to learn low-dimensional hidden variables that reflect the overall characteristics of the transient from the partial trends in the monitored data and can therefore use this vector as input for downstream diagnostic tasks, such as the identification of categorical or floating-point accident labels.

$$label_{class} = \mathrm{classifier}(\boldsymbol{latent})$$
$$= \mathrm{classifier}(\mathrm{encoder}(\mathcal{X}_n))$$
$$label_{scalar} = \mathrm{regressor}(\boldsymbol{latent}) \quad (8)$$
$$= \mathrm{regressor}(\mathrm{encoder}(\mathcal{X}_n))$$

## 3. EXPERIMENT AND ANALYSIS

In this section, we evaluate the proposed method through the LOCA diagnostic task of the HPR1000. First, the data generation method, the evaluation metrics, and the experimental details will be briefly described; then, some benchmark models will be used for comparison; finally, the experimental results will be analyzed and discussed.

### 3.1 Dataset Acquisition

The training data used in data-based reactor accident diagnosis models are generally generated from model-based system analysis programs or full-scale reactor simulators. The data set used in this paper is simulated by Advanced Reactor System Analysis Code, known as ARSAC, developed by the Nuclear Power Institute of China. ARSAC is based on a vapor-liquid two-phase non-uniform flow and non-equilibrium fluid dynamics model to solve heat transfer problems for vapor-liquid two-phase flows with non-condensable gases in a non-equilibrium thermal state [7]. A large number of general component models and special process models are built into the program for the construction of simple or various complex system loops. It covers the thermal-hydraulic transients and accident spectra of all nuclear power plants and has a powerful calculation capability and scope of application. Compared to other leading international system analysis software, such as CATHARE GB, WCOBRA/TRAC and S-RELAP5, the main features of ARSAC are: 1) Advanced matrix solving algorithm NRLU based on RCM (Reverse Cuthill-Mckee) chunk rearrangement technique; 2) A more refined wall heat transfer models; 3) A more refined re-inundation analysis module; 4) Advanced physical analysis module IAPWS-97.

Similar to RELAP, ARSAC constructs complex systems by writing input cards that define generic components and their connections. In this paper, the system boundaries for transient processes are modified on the basis of the HPR1000 steady-state card to form a transient card for calculating LOCA. In order to obtain the transient response of the system for different break locations and break sizes in the LOCA, after 500 s of steady state calculation, a break of different sizes in the cold or hot leg is inserted as the initiating event and 38 parameters in one loop are obtained as monitoring objects according to the instrumentation inspection system catalogue and the parameters are sampled at a frequency of twice per second for 100 s. A total of 346 LOCA accident transients were obtained, ranging from small breaks of 0.1cm to large breaks of 35.5cm in size, and including cold and hot leg breaks.

### 3.2 Optimization Objectives and Evaluation Metrics

The objectives are divided into optimization objectives for the autoencoder part and optimization objectives for the application of hidden variables to downstream diagnostic tasks.

In the autoencoder part, the most critical task is the ability to obtain an effective representation of transients based on the ability to fill in missing data and reduce noise. In this section, the degree of suppression of noise and the ability of the model to fill in missing data based on different ratios of noise and missing data will be evaluated. In order for the encoder-derived hidden variables to characterize the anomalous transients, it is necessary to make the reconstructed data as close as possible to the original, noise-free, non-missing monitoring data. The MSE is used here as the optimization objective for the network parameters, and is expressed as follows

$$\ell(x,y) = \{l_1,\ldots,l_N\}^\top, \quad l_n = (\mathcal{X}_n - \mathcal{X}_{re,n})^2$$
$$\arg\min_{\theta_{PAE}} \ell(x,y;\theta_{PAE}) \tag{9}$$

where $N$ is the batch size; $\mathcal{X}$ is the original monitoring data without noise; $\mathcal{X}_{re}$ is the reconstructed monitoring data; $\theta_{PAE}$ is the parameters to be optimized in Padded Auto-Encoder. In addition, in order to understand the substitution capacity of the hidden variables compared to the original data, the relative spatial distribution of the samples will be visualized in low-dimensions using a manifold learning method based on t-SNE. t-SNE is an outstanding method of dimensionality reduction for non-linear mapping of high-dimensional data to a low-dimensional space, implemented in two steps[8]. In the first step, t-SNE constructs a probability distribution over the high-dimensional data such that if two data points are more similar, they have a higher joint probability. In the second step, t-SNE constructs a probability distribution in the low-dimensional space such that the probability distribution in the low-dimensional space has a low KL divergence from the probability distribution in the high-dimensional space. As a result, the distribution of data in the lower dimensional space reflects the distribution of data in the higher dimensional space in a good way.

In the downstream diagnosis task, optimization is targeted at classifiers and regressors that use low-dimensional features for diagnosis. The goal of the classifier and regressor is to better 1) determine the location of the break and 2) calculate the size of the break in the LOCA transient, respectively. Since the location of the break is a categorial label, as for an MLP classifier, a cross-entropy loss function is used as the optimization objective

$$\ell_{cl}(x,y) = \text{sum}\{l_1,\ldots,l_N\},$$
$$l_n = -\sum_{c=1}^{C} \log \frac{\exp(x_{n,c})}{\exp\left(\sum_{i=1}^{C} x_{n,i}\right)} y_{n,c} \tag{10}$$

where $C$ is the number of categories, and $C=2$ in the identification of hot and cold leg breaks; $x_{n,c}$ is the predicted logical value for category $c$; $y_n$ is the one-hot code vector of data category labels; and $N$ is the batch-size. The size of the break is a numerical variable, so the mean square error is used as the objective loss function for the optimization

$$\ell_{re}(x,y) = \text{sum}\{l_1,\ldots,l_N\}, \quad l_n = (x_n - y_n)^2 \tag{11}$$

where $x_n$ is the size of the break predicted by the regressor; $y_n$ is the actual size of the break as a data label; and $N$ is the batch size. Thus, the joint loss function as the optimization objective is

$$\ell = \alpha \ell_{cl} + (1-\alpha)\ell_{re} \tag{12}$$

where $\alpha$ is the attentional weight of the loss function, chosen $\alpha = 0.5$ here because the break location and the size of the break are almost equally important in diagnosis.

In terms of metrics for diagnostic task, for the classification problem of determining the location of the break, we will use the Mcro-F1 of the classification result as the evaluation metrics; for the regression problem of predicting the size of the break, we use the Root Mean Squared Error (RMSE) as the evaluation metrics. The evaluation metrics are calculated as

$$Macro\text{-}F1 = \frac{1}{C}\sum_{c=1}^{C} F1_c \quad (13)$$

$$RMSE = \sqrt{\frac{\sum_{t=1}^{T}(\hat{y}_t - y_t)^2}{T}} \quad (14)$$

where $F1_c$ is the $F1$ value for category $c$; $C$ is the number of categories; $\hat{y}_t$ is the predicted break size; $y_t$ is the real break size label; and $T$ is the number of samples in the test dataset.

### 3.3 Experiment Conditions

During the construction of the autoencoder, since there are a total of 200 sampling points for each parameter, the size of each patch block is set to 40, and the length of the patched word vector sequence is $N=190$, and the word vector dimension is $D=40$. The dimensionality of the transient reactor feature vector after encoding is $\boldsymbol{latent} \in \mathbb{R}^{128\times 1}$. The number of Transformer layers in the encoder and decoder parts are $depth_{enc} = depth_{dec} = 4$. The number of heads in the encoder and decoder parts of the multi-headed self-attention is $heads_{enc}=heads_{dec}=4$. Dimensionality reduction ratio of intermediate hidden layers of feedforward neural networks is $ratio_{mlp}=0.8$. The dropout ratio for the fully connected layer is set to 0.1 to improve the overfitting of the network as a means of regularization.

For autoencoder and downstream diagnostic models built using deep learning models, the optimization method used is gradient descent, and the Adam algorithm with the addition of Nesterov Momentum is selected as the optimization tool[9]. It has a more flexible dynamic learning rate, which can improve the problem of falling into local minima in the process of parameter optimization. For the selection of the optimizer parameters, the initial learning rate is set to $lr = 1.0\times 10^{-3}$, the coefficients used to calculate the running average of the gradient and its square are $\beta_1 = 0.9$ and $\beta_2 = 0.999$, the smoothing coefficient is $\epsilon = 1.0\times 10^{-8}$, and the momentum decay coefficient is $4.0\times 10^{-3}$.

In each training epoch, for the same raw monitoring data, the autoencoder needs to learn separately for interference data with different signal-to-noise ratios and missing ratios. In the setting of the learning method, the autoencoder first learns the interference level in order from high to low, and then repeats the learning for the interference level of focus. The benefits of this training method are: 1) the autoencoder first learns for data with high interference levels during which the parameters are updated faster due to larger reconstruction errors; 2) when reconstructing data with low interference levels, the parameters can be updated more gently due to smaller reconstruction errors; and 3) finally, when training for specific interference levels, the parameters can be optimized more purposefully for the actual scenarios in which they may be applied. In the work of this paper, within each iteration step, the noise signal-to-noise ratio is added in the order of [20.0, 30.0, 40.0, 35.0, 35.0] and the data masking is added in the order of [0.40, 0.25, 0.10, 0.20, 0.20].

During the training process, the convergence is slower and the loss function fluctuates locally due to the stochastic masking process. The 1000th iteration is set as the time to stop training. The GPU device used for training was a single GT730 with 2GB of video memory. The total training time was 154.6 hours.

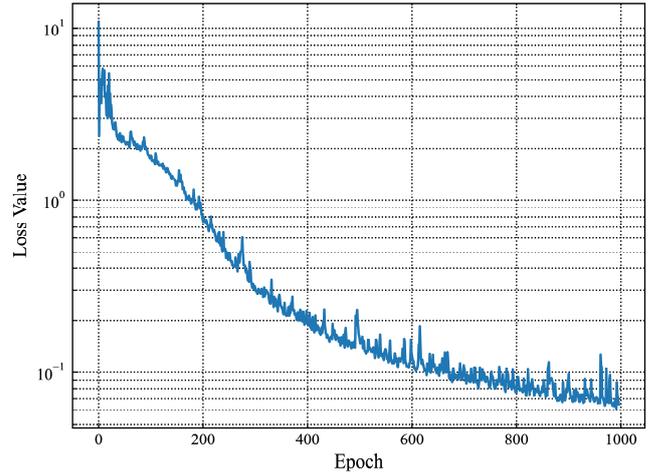

**Figure 2 Loss function descent during training autoencoder**

### 3.4 Auto-encoder performance

In order to demonstrate that the pre-trained model is able to effectively restore noisy and missing monitoring data information, we selected parameters that are vital to the monitoring for visualization. Hence, pressurizer pressure, pressure vessel water level and hot leg water level are used as example.

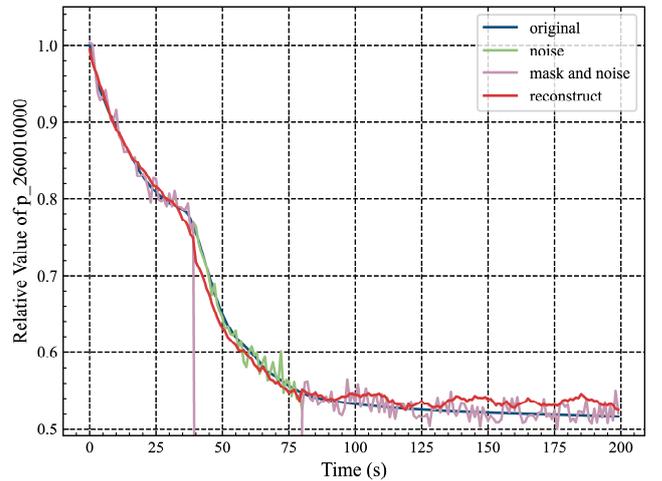

**Figure 3 Pressurizer pressure (SNR of 35, padding ratio of 0.2)**

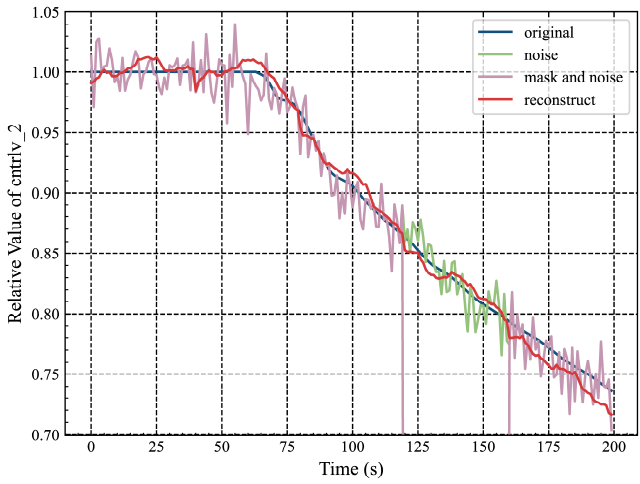

**Figure 4 PV water level (SNR of 35, padding ratio of 0.2)**

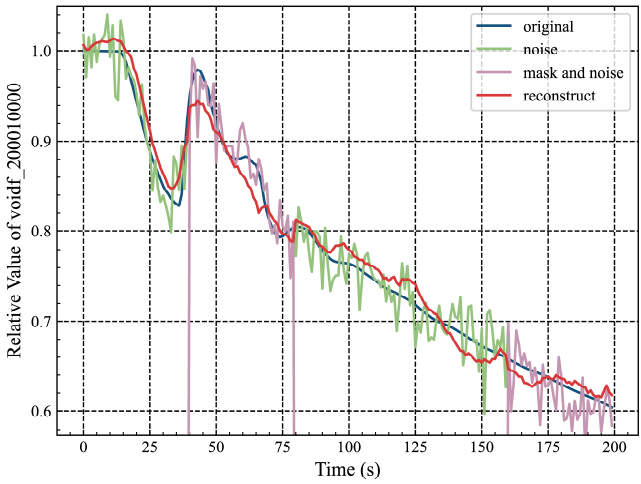

**Figure 5 Hot leg water level (SNR of 35, padding ratio of 0.2)**

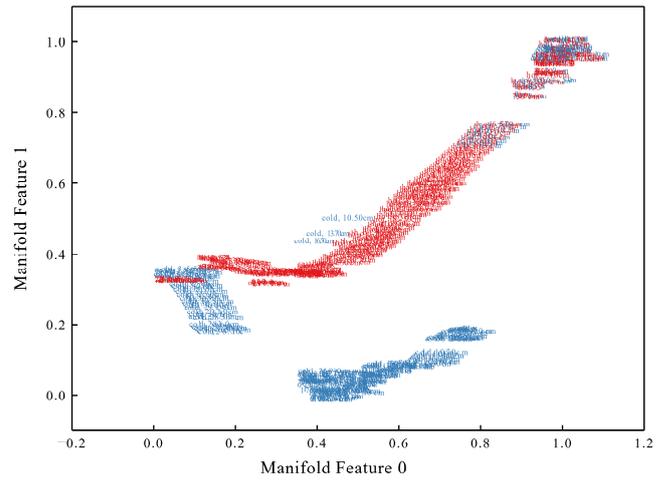

**Figure 6 Distribution of original noiseless and non-missing monitoring data**

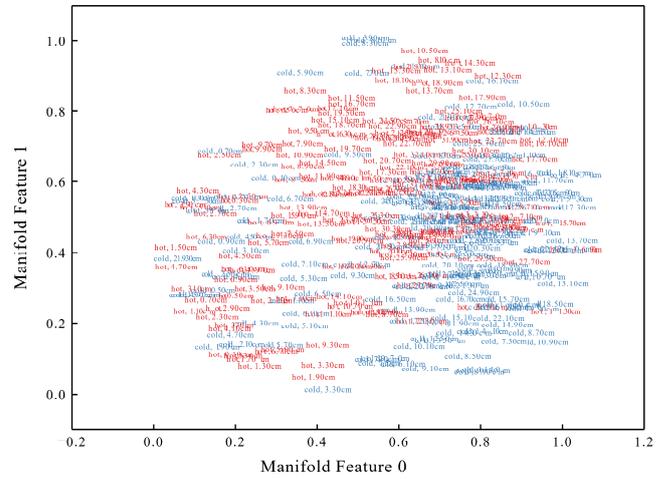

**Figure 7 Distribution of noisy and missing monitoring data**

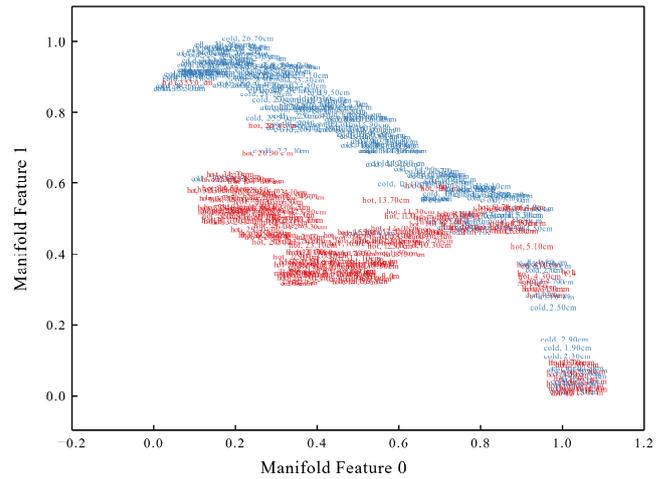

**Figure 8 Distribution of extracted features**

As shown in the Figure 3, Figure 4 and Figure 5, although noise is added to the monitoring data and some of the data are missing, the pre-trained autoencoder is able to fill in the missing parts of the monitoring data and suppress the noise to a certain extent. To ensure that effective features can be extracted using the encoder section of the autoencoder and can replace the original monitoring data, the spatial distribution of the high-dimensional data is mapped to two dimensions by the t-SNE manifold learning algorithm, which observes the distribution of the three datasets. The three datasets are 1) the original noiseless and non-missing monitoring data, as shown in Figure 6; 2) the noisy and missing monitoring data, as shown in Figure 7; and 3) the features extracted from the noisy and missing monitoring data, as shown in Figure 8.

Since the visualization of high-dimensional data using t-SNE manifold learning suffers from the randomness of the learning process, attention is mainly focused on the relative positions of individual data points rather than their absolute positions in the graph. It can be seen that the distribution of the data in the high-dimensional space is already a chaotic mess after noise addition and partial padding by zeros. But after extracting features from it using the encoder of Padded Auto-Encoder, the data features reappear as clusters and the relative distribution between data points has similar characteristics to the smooth and complete data.

### 3.5 Diagnosis Performance

The goal of this paper is to form input data that is useful for downstream diagnostic tasks by extracting valid features from monitoring data, so two types of diagnostic methods are compared: 1) the low-dimensional representations extracted from monitoring data containing noise and missing data are first used as input data, and then the diagnosis of break location and break size is performed, as shown in Figure 9(a); 2) the original monitoring data subjected to noise addition and partial missing are directly used as input to the diagnosis model, as shown in Figure 9(b). In the first class of methods, the performance of different classifiers and regressors will be compared, including SVM[10], Random Forest[11], XGBoost[12] and MLP. In the second type of methods, the previously proposed TRES-CNN method by the author, as well as the end-to-end models proposed by previous researchers, will be used as classifiers and regressors, and the diagnostic performance will be compared with the performance of the first type of methods.

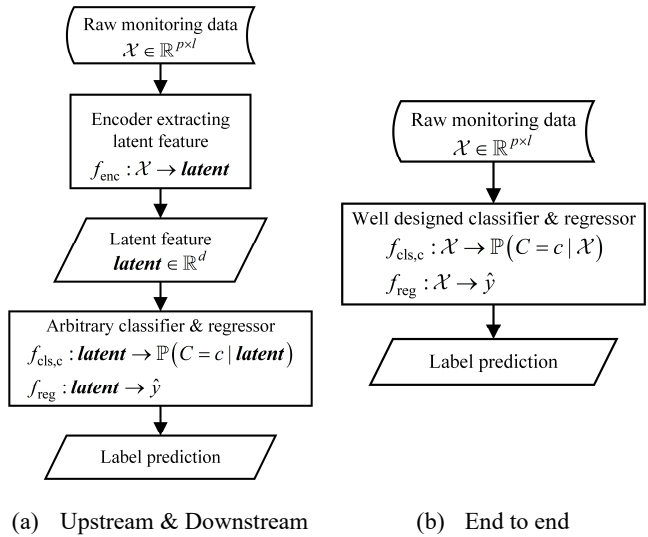

(a) Upstream & Downstream     (b) End to end

**Figure 9 Differentiated diagnostic framework**

The diagnostic experiment uses monitoring data containing noise with a ratio of 0.1 missing and SNR of 35. If a multi-layer perceptron containing three hidden layers and one output layer is used as the classifier and regressor after feature extraction, and Eq. (12) is used as the loss function for optimization, then after 256 iterations, the precision of cold leg breakage diagnosis is 90.5%, and the precision of hot leg breakage is 93.4%, with $Macro\text{-}F1=0.919$ and $RMSE=0.322$. The experimental results compared to other diagnostic methods are shown in Table 1.

**Table 1 Diagnosis performance of different methods**

| Type | Method | Cold leg precision | Hot leg precision | Macro-F1 | RMSE |
|---|---|---|---|---|---|
| Upstream & Downstream | MLP | 0.905 | 0.934 | 0.919 | **0.312** |
| | SVC/SVR | 0.898 | 0.917 | 0.908 | 0.649 |
| | XGBoost | **0.911** | **0.946** | **0.927** | 0.564 |
| | Random Forest | 0.858 | 0.871 | 0.864 | 0.982 |
| End to end | TRES-CNN | 0.682 | 0.646 | 0.662 | 2.945 |
| | BPNN[13] | 0.612 | 0.588 | 0.598 | 3.623 |
| | CNN [14] | 0.641 | 0.675 | 0.656 | 3.259 |

As the data in the Table 1 shows, the diagnosis method by feature extraction followed by downstream diagnosis has a significant improvement compared to the direct end-to-end learning method. In particular, the use of end-to-end diagnostic models shows significant distortion in the case of interference from noise and missing data. At the same time, in since the dimensionality of the extracted effective features is already low enough, the use of classical statistical learning models, such as SVM and XGBoost, is also able to achieve similar or better performance than connectionist-centered deep learning algorithms. This means that with a suitable pre-trained feature extraction model, it is possible to build effective and robust accident diagnosis models for noisy and missing data at the user side at a lower cost.

### 4. CONCLUSIONS

In this paper, a Padded Auto-Encoder model is proposed first. The autoencoder not only learns the effective low-dimensional features of the model during the compression and reduction of the data, but also complements the missing data and reduces the noise level. Then, a method is proposed to perform diagnosis using the encoder part of the pre-trained autoencoder. This method first uses the encoder to extract the effective low-dimensional features of the original data disturbed by noise and missing data, and then uses a simple learning algorithm to classify and regress the features. This two-step approach to diagnosis has significant benefits over direct end-to-end diagnosis methods, receiving an improvement of 41.62% and 80.86% in the classification and regression task metrics respectively, in that the extracted features are more robust to the diagnosis of disturbed data, and the training cost of downstream classifiers and regressors is reduced.

### REFERENCES

[1] Zhao, X., Kim, J., Warns, K., Wang, X., Ramuhalli, P., Cetiner, S., Kang, H. G., and Golay, M., 2021, "Prognostics and Health Management in Nuclear Power Plants: An Updated Method-Centric Review With Special Focus on Data-Driven Methods," Frontiers in Energy Research, **9**, p. 294.

[2] Li, X., Fu, X.-M., Xiong, F.-R., and Bai, X.-M., 2020, "Deep Learning-Based Unsupervised Representation Clustering Methodology for Automatic Nuclear Reactor Operating Transient Identification," Knowledge-Based Systems, **204**, p. 106178.

[3] Kim, H., Arigi, A. M., and Kim, J., 2021, "Development of a Diagnostic Algorithm for Abnormal Situations Using Long Short-Term Memory and Variational

Autoencoder," Annals of Nuclear Energy, **153**, p. 108077.

[4] Naito, S., Taguchi, Y., Kato, Y., Nakata, K., Miyake, R., Nagura, I., Tominaga, S., and Aoki, T., 2021, "Anomaly Sign Detection by Monitoring Thousands of Process Values Using a Two-Stage Autoencoder," Mechanical Engineering Journal, **8**(4), pp. 20-00534-20–00534.

[5] Dosovitskiy, A., Beyer, L., Kolesnikov, A., Weissenborn, D., Zhai, X., Unterthiner, T., Dehghani, M., Minderer, M., Heigold, G., Gelly, S., Uszkoreit, J., and Houlsby, N., 2021, *An Image Is Worth 16x16 Words: Transformers for Image Recognition at Scale*, arXiv:2010.11929, arXiv.

[6] He, K., Chen, X., Xie, S., Li, Y., Dollár, P., and Girshick, R., 2021, "Masked Autoencoders Are Scalable Vision Learners," arXiv:2111.06377 [cs].

[7] Zhou, J., Wu, D., Ding, S., and Jiang, G., 2020, "Research on Analysis and Modeling Methods for LBLOCA in Nuclear Power Plants Based on Autonomous LOCA Analysis Platform ARSAC."

[8] Van der Maaten, L., and Hinton, G., 2008, "Visualizing Data Using T-SNE.," Journal of machine learning research, **9**(11).

[9] Dozat, T., 2016, "Incorporating Nesterov Momentum into Adam."

[10] Cortes, C., and Vapnik, V., 1995, "Support-Vector Networks," Mach Learn, **20**(3), pp. 273–297.

[11] Breiman, L., 2001, "Random Forests," Machine Learning, **45**(1), pp. 5–32.

[12] Chen, T., and Guestrin, C., 2016, "XGBoost: A Scalable Tree Boosting System," *Proceedings of the 22nd ACM SIGKDD International Conference on Knowledge Discovery and Data Mining*, Association for Computing Machinery, New York, NY, USA, pp. 785–794.

[13] Basu, A., and Bartlett, E. B., 1994, "Detecting Faults in a Nuclear Power Plant by Using Dynamic Node Architecture Artificial Neural Networks," Nuclear Science and Engineering, **116**(4), pp. 313–325.

[14] Lee, G., Lee, S. J., and Lee, C., 2021, "A Convolutional Neural Network Model for Abnormality Diagnosis in a Nuclear Power Plant," **99**, p. 106874.

**AUTHOR'S INFORMATION**

Chengyuan Li is a master's student at NPIC with research interests in artificial intelligence algorithm-based reactor safety analysis and can be reached at email address lichengyuan98@gmail.com.